\documentclass[12pt,a4paper,final]{article}

\usepackage{wasysym}
\usepackage{graphicx,booktabs,geometry}
\usepackage{url}

\usepackage{natbib}
\usepackage[resetlabels,labeled]{multibib}
\newcites{Sample}{Articles in the study sample}

\title{Moving beyond harm. A critical review of how NLP research approaches discrimination}
\date{2025\\ January}
\author{Katrin Schulz, ILLC, University of Amsterdam
\and Marjolein Lanzing, Philosophy Department, University of Amsterdam
\and Giulia Martinez Brenner, University of Amsterdam}

\begin{document}

\maketitle

\begin{abstract}
How to avoid discrimination in the context of NLP technology is one of the major challenges in the field. We propose that a different and more substantiated framing of the problem could help to find more productive approaches. In the first part of the paper we report on a case study: a qualitative review on papers published in the ACL anthologies 2022 on discriminatory behavior of NLP systems. We find that the field (i) still has a strong focus on a technological fix of algorithmic discrimination, and (ii) is struggling with a firm grounding of their ethical or normative vocabulary. Furthermore, this vocabulary is very limited, focusing mostly on the terms "harm" and "bias". In the second part of the paper we argue that addressing the latter problems might help with the former. The understanding of algorithmic discrimination as a technological problem is reflected in and reproduced by the vocabulary in use. The notions of "harm" and "bias" implicate a narrow framing of the issue of discrimination as one of the system-user interface. We argue that instead of "harm" the debate should make "injustice" the key notion. This would force us to understand the problem of algorithmic discrimination as a systemic problem. Thereby, it would broaden our perspective on the complex interactions that make NLP technology participate in discrimination. With that gain in perspective we can consider new angles for solutions.
\end{abstract}

\section{Introduction}

\noindent In recent years a lot of attention has been paid to machine learning applications that show unfair or biased behavior towards particular groups or individuals. This led to growing societal concerns about harmful discrimination and reproduction of structural inequalities once these technologies become institutionalized in society. Identifying and resolving algorithmic bias and achieving algorithmic fairness has therefore become an important research field. With the rise of large language models (LLMs) and the speed with which they are integrated in all aspects of life, this has been particularly true in the area of Natural Language Processing (NLP).

However, it has been argued that in AI research and policy, the remedies against discriminatory algorithms are often narrowly framed as addressing technological design problems aimed at compliance, rather than complex, structural, social-political issues. This leads to a highly techno-centric and technocratic approach towards algorithmic fairness \cite{balayn}. Furthermore, it has been argued that this framing stands in the way of successfully addressing the problem of negative societal impact \cite{blodgett_language_2020,birhane_algorithmic_2021}. 

In this paper we want to engage with this line of argument. Firstly, we will investigate whether and in what sense it still holds that algorithmic discrimination is approached in a techno-centric way. To this end, we perform a critical interpretative literature review, drawing a sample from the publications in the ACL anthologies for the year 2022 (see Section~\ref{sec:methodology} for a description of the methodology). We find that nearly all the work on the topic of algorithmic discrimination in our sample still strongly focuses on finding technological solutions for the problem (see Subsection~\ref{subsec:rq3}). We also observe that this comes together with a rather narrow framing of the mechanisms underlying the problem (Subsection~\ref{subsec:rq2}) and a weak normative/ethical grounding of the problem addressed (Subsection~\ref{subsec:rq1}).

Building on these results, in Section~\ref{sec:reframing} we provide a proposal for how to overcome these limitations. The driving idea of this proposal is that in order to see the full range of factors responsible for algorithmic discrimination, and, thus, move beyond a technological solution, we need a thorough understanding of the problem. The  proposal, therefore, targets the particular way of how the field views and conceptualizes the problem of algorithmic discrimination. Concretely, we will criticize the language used to develop the normative grounding of NLP research on algorithmic discrimination. As our case study shows, the current debate centers around the notions of {\it bias} and {\it harm}. We will argue that (i) {\it bias} in itself cannot provide ethical grounding and (ii) {\it harm} is neither sufficient nor necessary for discrimination to take place.

The alternative framing that we propose centers around the notions of {\it structural and systemic injustice} \cite{haslanger22,haslanger23, haslanger24}. Focusing on structural and systemic injustice instead of individual harms creates space to consider more complex mechanisms underlying patterns of algorithmic discrimination. It also changes the type of mechanisms we are looking for: a structural and systematic injustice approach would reveal unjust social-political structures connected to ideologies of oppression as opposed to individual actions of (AI) agents or particular design choices made in the development of this technology. We hope that such a re-framing of the problem will help to create awareness among researchers and engineers about alternative ways to approach developing responsible NLP technology, but will also lead to a broader engagement of the academic and non-academic community in the discussion of how to address ethical issues of NLP technology.

Engaging with ethical questions can be complex and challenging. Before we continue with the main body of the paper, let us be explicit about what kind of unethical behavior of NLP technology we are targeting here. By restricting our discussion to algorithmic discrimination, we are looking specifically at the involvement of language technology in practices that lead to the unjust exclusion or mistreatment of groups or individuals. This still includes a broad spectrum of ways in which NLP technology can be involved in unethical practices, but not necessarily all such practices. It does include problems often discussed using notions like discrimination, fairness and bias, among others. 

Second, making claims about unjust practices involves taking a normative stance. There is a lot of discussion about what morally wrong behavior is and we do not want to build this work on one very specific approach to ethics and, thereby, limit the reach of our proposal. However, drawing some bottom line, making some basic assumptions about what we value is unavoidable when grounding a normative position. For this paper our baseline is democracy. This will be the foundation on which we build our normative claims and arguments. There is a very lively debate in political science and philosophy about what democracy exactly entails and how it should be implemented.\footnote{For an overview of the philosophical debate see \cite{sep-democracy}. A central publication in the field of political science is \cite{dahl2020democracy}.} To make our argumentation as broadly applicable as possible, we choose here a very general and thin concept of democracy. Let us agree that this democracy encompasses a liberal state of justice \cite{rawls}. This includes valuing individual (human) rights, without failing to take into account the importance to recognize collectives and differences in order to grasp structural and systemic injustices that escape the individualist eye \cite{Young}. The underlying democratic ideal is that all members and dwellers of a democratic society should have (access to) equal rights and equal opportunities to participate, including the basic conditions necessary for this \cite{rawls}. Building on this, the normative question of what we should expect of technology -- which lies at the core of this paper -- comes down to asking whether technology supports or hinders our democratic values and the processes that we have put in place to foster equality and freedom for all members of society. First of all, this grounds the normative stance that technology, at the bare minimum, should not reinforce oppression through unjust structures and practices of discrimination because it should not violate the principles of democratic equality. It also provides a baseline from which we can evaluate solutions proposed to prevent language technology to participate in discriminatory practices.
\newline
\newline



\section{The Case Study - Methodology}\label{sec:methodology}

As explained in the introduction, our first goal was to investigate whether and in what sense there is indeed a focus on a techno-fix when it comes to addressing discriminatory behavior of NLP technology. In order to answer these questions we performed a critical interpretative literature review. This type of review integrates the methods used in systematic reviews with a qualitative tradition of inquiry. It includes a transparent methodology for how sources of information are selected, but it does not aim to include all relevant sources \cite{Palmer}. For our purposes a full coverage of the literature is not necessary, because we are looking for general developments in the methodology applied in NLP research. 

The assessment of the literature was performed in two steps. First, we needed to filter out articles that are not on our target topic. Then, we qualitatively assessed the selected articles with respect to our research questions. 

\paragraph{Step 1: Filtering}

For our case study we used the ACL anthologies\footnote{https://aclanthology.org/} and focused on all papers published in  2022. This is a choice that can be debated. Choosing ACL is in itself already sampling the publications on language technology. Not all the venues and journals publishing relevant research are included. The question is whether this choice is biased and affects our conclusions. We will come back to this in the discussion section~\ref{sec:discussion}. Because of the rapid developments in NLP research, we decided to focus on one year in particular, as this approach may facilitate later studies examining the field's evolution over time.\footnote{Our results can be compared to the work of \cite{blodgett_language_2020} that studies similar questions, but only looking at publications up to 2019.} From the 84.000 ACL publications from 2022, we step-wise filtered out all publications that studied specifically the role of NLP systems in discriminatory practices. We filtered all titles and abstracts using a list of keywords. For this we selected a list of words used in the context of describing discrimination (see Table~\ref{tbl:keywords}). After removing duplicates, a list of 646 papers remained. Two encoders read the abstracts of these papers to determine whether the paper was really addressing unethical behavior of an NLP system. If the annotators disagreed or couldn't decide based on the abstract whether discrimination were emphasized, they continued reading the introduction, and discussed the article, until an unanimous decision was reached. 

The filtering process resulted in a corpus of 207 ACL publications. From this corpus we took a random sample of 50 papers which were then coded with respect to the research questions.\footnote{Some of the papers selected did not fit our criteria when we looked at the body of the papers. These papers where replaced by other random choices from the database. In the end, we replace 14 papers from the initial sample for this reason. We will come back to this in the discussion.}\label{wrong_annotation}

\begin{figure}
\begin{tabular}{l|p{5,5cm}}
{\bf keywords} & “stereotype”, “stereotypes", “bias”, “fairness”, “injustice”, “discrimination”, “disparity”, “disparities", “equality”, ”equity”.
\end{tabular}
\caption{Keywords used for filtering the full body of papers.}
\label{tbl:keywords}
\end{figure}

\paragraph{Step 2: Coding}
The purpose of this case study was to investigate whether there is indeed a focus on a techno-fix in NLP research on unethical behavior of AI. To answer this question we needed to map rather generally how researchers engage with the problem of discriminatory NLP technology. In order to do so we split our main research question into three sub-questions.

\begin{enumerate}
    \item How do the authors describe the problematic behavior of the technology they are addressing?
    \item What are the causes or mechanisms that are given to underlie the problematic behavior of the technology?
    \item What solutions are proposed?
\end{enumerate}

The first question targets the framing of the problem. We were interested in seeing whether the authors describe the issue as an ethical issue and what kind of ethical terminology they use to do so. The concern here was that a primary focus on the technological aspects of the problem might lead to a neglect of the ethical and conceptual grounding of the work. The second question aims to map what the authors consider to be possible causes of algorithmic discrimination. These causal dependencies are a good indicator of the solution space they are considering. If algorithmic discrimination is framed as a technological problem, we should find predominantly technological causes mentioned here. The final question targets the solutions that are actually discussed in the papers. A focus on a tech-fix should be clearly visible in the type of solutions here given.

We then formulated label groups for each of these sub-questions (for an overview of the labelss see the Appendix). For the first question we defined three label groups: (i) do the authors make a normative/ethical statement, (ii) do the authors provide an explanation for the normative/ethical claim made, and (iii) is the explanation grounded in the theoretical literature. For the second questions, we started out labeling the type of cause for the wrongful behavior of the AI system. Because of the strong focus on bias in an AI system as cause, we also additionally labeled the causes given for this bias. The rationale behind this choice was that this would allow us to trace back the mechanisms taken to be responsible for the problematic behavior of the technology through multiple steps. For the last question, we labeled the type of solution for all solutions discussed in the paper. Additionally, we marked the solution that the authors selected to work on.

We used ATLAS.ti was used for the coding of the articles. For the four main label groups {\it ethical grounding, cause of wrongful behavior, cause of bias} and {\it solution} we started with a very fine-grained system of labeling and then merged labels when patterns started to emerge. The full set of labels can be found in the Appendix.

Two encoders read the introduction, the discussion, the conclusions and any other passages of the paper that contained occurrences of the (see Table~\ref{tbl:keywords}).



\section{The case study -- Results}
\label{sec:results}

\subsection{Sub-question 1: conceptual and normative grounding}
\label{subsec:rq1}

When looking at how scientists describe the problematic behavior of NLP systems we could distinguish four groups of descriptions (for examples see the appendix):

\begin{itemize}
    \item[(i)] The system behavior that the paper addressed was never described as negative or unwanted. These were papers that discussed bias measurement/analysis in NLP systems without stating that bias could potentially be problematic.
    \item[(ii)] The system behavior the paper addressed was described as negative, but not explicitly as ethically wrong. In this case either (a) terms such as "unwanted" or "negative" were used without clarifying that the behavior was morally problematic, or (b) the discussion focused on legal violations rather than ethical considerations,\footnote{We gathered reference to law violations under this label, because it concerns adherence to a different type of norm: you can avoid doing something because it is against the law, without that you have to believe that it is morally wrong. Of the three papers with that label two concern privacy issues. In both cases it was not entirely clear whether privacy was clearly seen as an ethical or only a legal issue.} or (c) the authors referred to issues commonly associated with unethical AI — such as unsafe systems, lack of trustworthiness, or insufficient transparency — without framing these concerns explicitly in terms of moral wrongdoing. 
    \item[(iii)] The negative impact of technology was described in terms of harm.
    \item[(iv)] Other ethical vocabulary was used to describe the negative impact.
\end{itemize}

\begin{figure*}[t]
\centering
\includegraphics[width=1\textwidth]{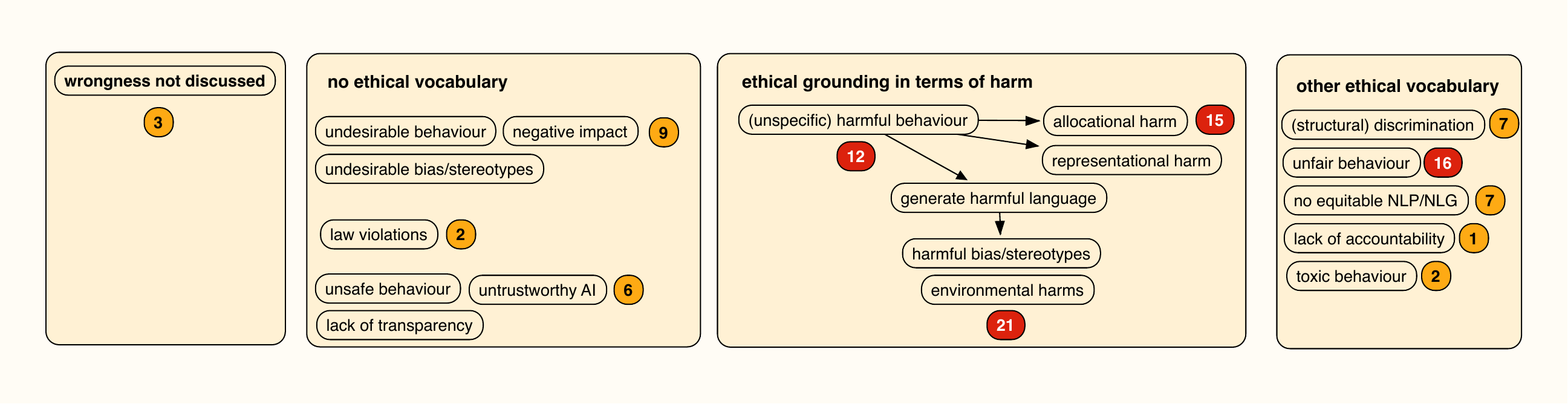} 
\caption{The figure displays the labels used to code ethical grounding. From left to right the ethical vocabulary becomes more developed. The numbers next to a label group reports the number of papers in which this label/label group is used. The red numbers represent the dominant labels/label groups in the sample. The arrow represents label-dependence. A single paper could be labeled with different quote-level labels.}
\label{fig:norm}
\end{figure*}

An overview of the labels in each group, together with the counts of papers in which the label occurred is given in  Figure~\ref{fig:norm}. There is a strong dominance of framing the problematic behavior in terms of harm. The harm concerned nearly exclusively harm caused by the output of the system. A single paper in the sample referred to harm that can occur during production of AI systems ([Sample13]
). Apart from harm, also framing in terms of fairness was frequent. Other related vocabulary is also used but much less frequent. Furthermore, we noticed that there is still as substantial group of papers that do not frame the problem as ethically wrong, but instead as undesired or make no clear normative statements.

We also often see that a paper makes a normative claim, but without any additional explanation about why the behavior is wrong, to whom and in what situation. An example is [Sample4] 
, a paper that discusses measures of social biases in prompt-based Multi-Task learning. The only normative statement in the Introduction section occurs in the description of the contribution of the paper: ``BBNLI measures human cognitive biases across three different domains and capturing 16 {\it harmful stereotypes}.'' (emphasis added by the authors). Nowhere in the paper is it explained what biases are, how they are related to stereotypes, and what is wrong about biases and stereotyping in general. From the fifty papers in the sample, only 22 papers attempt to explain at least once the normative/ethical claims made. 

There is also limited grounding in interdisciplinary literature on discrimination. Only twenty-five of the fifty papers referred to non-technologial work on the topic.\footnote{There were papers that did not provide an explanation for their normative statements, but did cite theoretical literature. For instance, see [Sample9] 
: ``Deploying such biased word embeddings in downstream tasks would cause allocational and representational harms (Blodgett et al., 2020).''.} Seven of these papers quoted \cite{blodgett_language_2020} in this context. We noticed that grounding occurred more frequently in case the wrong was framed in terms of fairness. This is probably the case because there exists a broader spectrum of accessible literature on the issue of fairness than for bias and harm. Also papers using the terms "allocational harm" or "representational harm" tended to be more explicit about their normative grounding. In this case a very clear reference exists (\cite{Barocas}) that can provide guidance.

Even in case a paper did offer some explanation for how or why the targeted behavior of an AI system was wrong, this explanation was in general not very well developed and the literature provided very limited. For instance, we labelled the quote ``... language models (LMs) that empower much of the web as we know it are well known to contain biases that promote structural discrimination in downstream tasks against minorities and larger social groups alike (Bender et al., 2021).'' [Sample23] 
as providing a normative explanation, because a link is made between biases and discrimination in downstream tasks. However, the paper does not define the concept of bias or explain what is meant by structural discrimination in downstream tasks, linking to the substantial philosophical and socio-theoretical work on the matter. Thus, the key concepts of the research are still poorly grounded.

\subsection{Sub-question 2: Causes of wrongful behavior}
\label{subsec:rq2}

In the next step, we looked at what causes were given in the papers for the unethical or unwanted behavior of language technology. We could distinguish four groups of root causes that researchers pointed to: (1) the training of the technology, (2) technological causes, (3) causes related to the research or development agenda and (4) causes related to the use of the technology (see the bold labels in Figure~\ref{fig:causes}; examples for all label groups can be found in the appendix). The cause stated most often offered was bias, either bias in the training data (issues with training) or biased behavior of the AI technology (technological causes). If researchers pointed to issues with the training data as the cause of the wrong behavior of the AI system, the wrong was described as bias in the system. As the reader can see, very few papers mentioned other causes. In most cases bias was stated as the cause, without that it was made clear what exactly was meant by bias and how this bias caused the unethical behavior.

\begin{figure*}[t]
\centering
\includegraphics[width=1\textwidth]{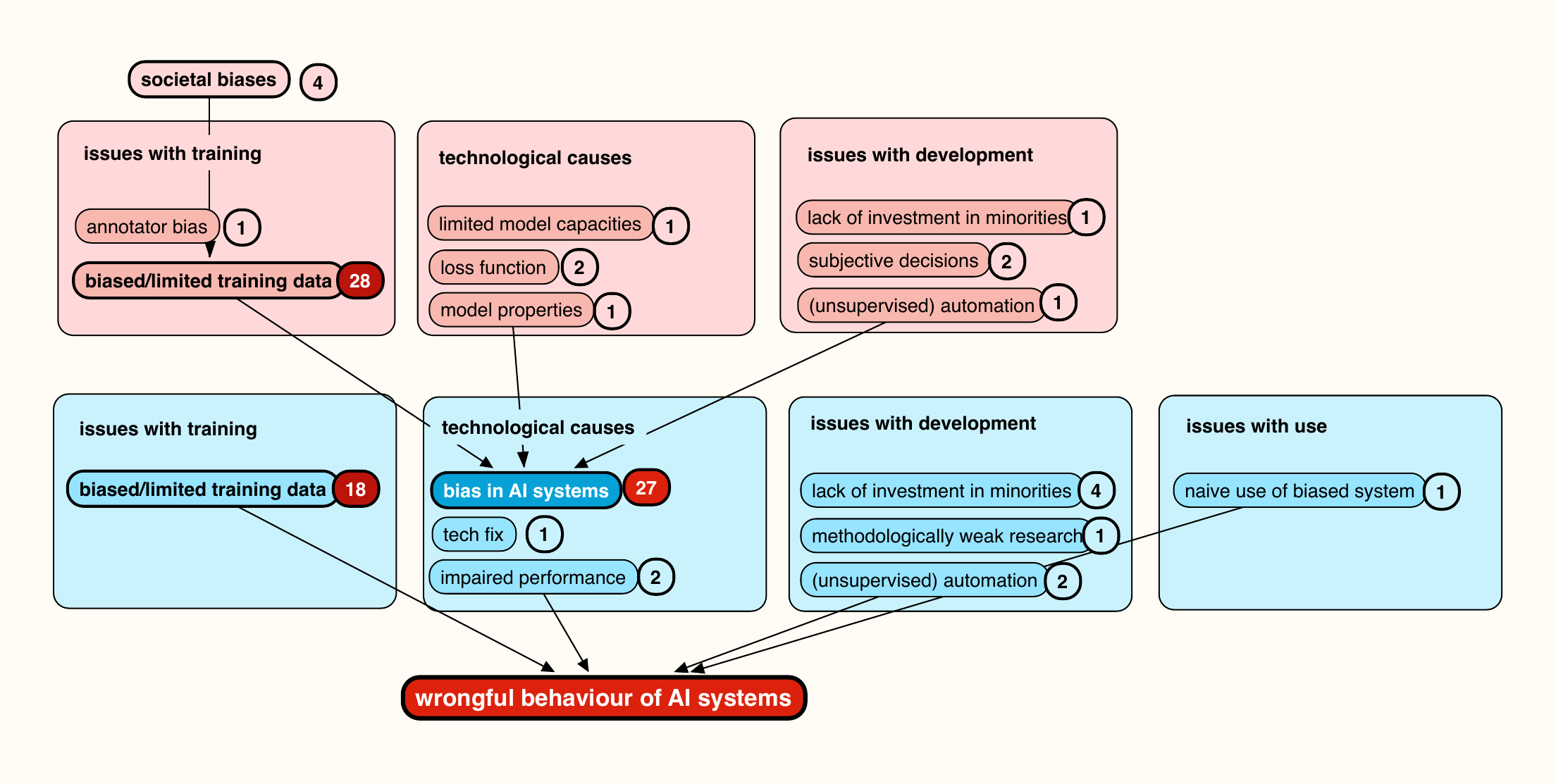} 
\caption{The figure displays the labels used to code the causes discussed in the papers for the wrongful behavior of NLP technology. The upper, read labels code causes provided for bias in AI systems, if this was listed as cause of the wrongful behavior. The numbers next to a label group give the number of papers in which this label is used. The red numbers represent the dominant labels in the sample. Arrows represent causal dependence}
\label{fig:causes}
\end{figure*}

We also labeled causes of bias in AI systems if the latter was mentioned as cause of the wrongful behavior, see the red boxes in Figure~\ref{fig:causes}. Here, most authors saw biased or limited training data as cause for the bias in the AI system, see, for instance [Sample40] 
: ``Disparities may reflect data imbalances: If training data contains less data from a group, predictions for that group will tend to be worse.'', or [Sample26] 
: ``... there is a growing concern if such language models contain social biases such as stereotyping negative generalizations of different social groups and communities, which might have been present in their training corpora \cite{liang2021towards, garrido2021survey}.''. Very few alternative causes were discussed. The same classification emerged as with the primary cause of the unethical behavior: the cause was either located in the training, or technological issues were blamed, or the research or development agenda. Issues with use were not relevant in case we look at bias in an AI system. 

In sum, the causal narrative that emerges is relatively straightforward and predominantly centered on the technology itself: the wrongdoing is attributed to a biased AI system, which in turn is biased due to limitations/bias in the training data. This framing subsequently shapes the space within which researchers seek solutions to the system's problematic behavior.

\subsection{Sub-question 3: Solutions proposed}
\label{subsec:rq3}

Lastly, we labeled all solutions mentioned by the authors of the articles. We, again, found four groups of solution types: 1) purely technological solutions, 2) solutions that addressed the behavior of the researcher/engineer, 3) solutions concerning institutional behavior, and 4) solution targeting the behavior of other agents. We found that technical solutions strongly dominated, as can been seen in Figure~\ref{fig:solutions} (again, examples for all label groups can be found in the Appendix). Additionally, nearly all the papers focused in their research only on technical solutions. We labeled papers that worked on bias measurement separately, even though these are also technological solutions, because bias measurement can also be pursued to raise more general awareness of the problem. However, in most cases, even if awareness was given as reason to develop bias measures, this was described as awareness of the researchers, and not those of the user or other agents. For example, in [Sample3] 
the focus is on awareness, but this is awareness of design choices when measuring bias in language generation. \footnote{[Sample24] 
appears to be an exception and targets general awareness of the limitations of language models.} 

\begin{figure*}[t]
\centering
\includegraphics[width=1\textwidth]{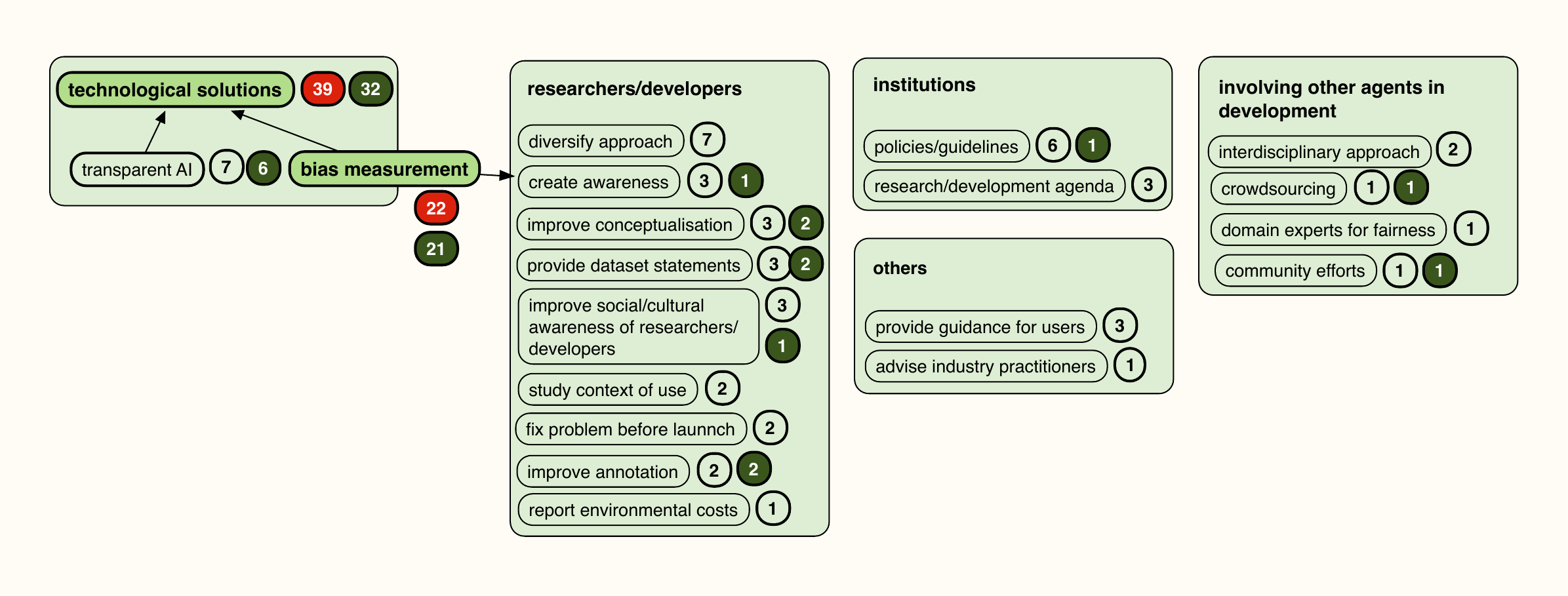} 
\caption{The figure displays the labels used to code solutions. The numbers next to a label report first the number of papers in which this solution is mentioned and, secondly, the number of papers in which this is the solution the paper works on. Papers can work on multiple solutions (for instance, measuring bias and mitigating bias). The red numbers represent the dominant labels.}
\label{fig:solutions}
\end{figure*}

There was only one paper in the sample that did not present a technological solution/bias measurement as its main contribution (i.e. [Sample40] 
, in which regulative measures for the employment and development of NLP technology are discussed\footnote{This also the single paper discussing discrimination through the production of NLP technology.}). Furthermore, the focus of nearly all solutions is on fixing the system. Improving annotation processes or involving other agents in development is in the end still for the purpose of building technology that solves the issue of algorithmic discrimination. This applies to nearly all the solutions proposed under the label groups {\it researchers/development}, {\it institutions}, and {\it involving other agents in development}. 


\subsection{Conclusion}

With respect to our first question we see from our results that the conceptual grounding of the normative claims made in the literature is still weak. While more than 2/3 of papers use ethical terms to describe the problematic behavior of the NLP technology they discuss, the description remains rather superficial and not well anchored the in theoretical literature. Concretely, in the majority of the papers nothing is said about why the system's behavior is harmful, to whom and in what context etc..

With respect to our second question concerning the causes discussed for the wrongful behavior of the AI systems we observe a rather limited and uniform view on the causes. Furthermore, there is a clear focus on locating the problem in the technology and blaming bias for it. We however found very little reflection on the interaction between technology and society or the mechanisms that determine developmental agendas or choices of datasets. 

The solutions discussed strongly focused on technical solutions. Most of them concerned bias measurements. Other solutions were considered, but rarely the focus of the work reported on. Many of those solutions also served the purpose of improving the technology.

In sum, our finding indicate that, in 2022, NLP-research on discrimination was strongly oriented toward technological solutions. The research landscape seems to form a techno-centric research bubble with a very clear consensus on what a relevant research questions are and how they should be addressed. The predominant focus was on identifying and mitigating bias in NLP systems using technological approaches. There was limited concern for the complex interaction between technology and society and the majority of studies lacked substantial theoretical grounding in socio-technical or ethical frameworks.



\section{Discussion}
\label{sec:discussion}

\subsection{Methodological limitations}

The validity of our results depends on the representativeness of our sample. There are large conferences addressing ethical and societal implications of AI technology like FAccT, AIES and EAAMO that are not included in the ACL anthology. However, these conferences are not specifically about NLP research, which is the topic of the present manuscript. This is why we opted for using ACL. But there will be papers addressing language technology that went to these conferences specifically because they focused on societal and ethical issues. 
In future work we will extend the search space for relevant papers. 

There are two steps in the selection of the sample from our database that could have biased the sample. First of all, we might have chosen the wrong set of keywords for the first filtering. The worry here would be that our set misses key terminology used in the debate of discrimination in NLP and, therefore, we filtered out part of the discourse. However, these terms would have come up during close reading of the texts, which did in fact happen for ``harm''. 
This was the only additional term that we found during close reading of the sample. We therefore think that the risk that we missed an important part of the discourse that would change our results is small.

The second error-sensitive step is the manual filtering and annotation of the sample. We already noticed in footnote~\ref{wrong_annotation} that we had a high rate of false positives in the set of papers selected as addressing discrimination. Although false positives were filtered out in later stages of the selection process and thus are unlikely to have significantly affected the results, their high number raises concerns about the potential presence of false negatives. However, we observed that most of the false positives, i.e. papers initially classified as relevant, mention the societal impact of the discussed technology in the abstract, but do not return to this point in the full text. This observation supports the assumption that if a paper genuinely engages with issues related to algorithmic discrimination, such relevance would likely be reflected in the abstract, given that societal impact is often used as a selling point. Therefore, we consider the likelihood of false negatives to be low.

\subsection{Putting the results in perspective}

If we agree that there is a strong focus on a techno fix in the research literature, we still do not need to agree that this is problematic. One might argue, for instance, that finding a focus on technological solutions is neither surprising, nor problematic given that we looked at a sample of papers written by NLP engineers and researchers. So, of course they will focus on technological solutions. 

Let us draw a distinction between the group of NLP practitioners and the field of NLP. NLP practitioners, as we understand them here, are the people who work on the technological part of NLP systems. These are the code developers, the data scientists and engineers that do the actual implementation and training of the systems. Apart from the practitioners, however, we have the {\it field} of NLP, which is the study of NLP technology, naturally in relation to the society that builds and employs this technology. The standards for what is done in the field of NLP should not be set by what NLP practitioners could do, but by what needs to be done to tackle the problems we face. Developing technological fixes for algorithmic discrimination is not the only, and arguably also not the best way to address the issue. Consequently, we should find work on such alternative solution in the field. Our case study suggests that this work is missing.


A lot needs to be invested into NLP research to make sure that we develop technology that is beneficial to society. But this involves more than just engineering. We are certainly not the first to point this out. In \cite{blodgett_language_2020}, for instance, very concrete proposals for research questions are made that need to be answered. However, NLP practitioners cannot answer these questions by themselves.\footnote{Just to give an example, the first of the questions proposed in \cite{blodgett_language_2020} is: ``How do communities become aware of NLP systems? Do they resist them, and if so, how?''. This is a question that requires research methodology that is not in the toolbox of an NLP practitioner.} This needs to be an interdisciplinary effort, in dialogue with society. 


This also answers a second potential objection: Why not adopt a techno-centric approach to algorithmic discrimination as long as it solves the problem? The issue with this objection is that despite an immense investment of research in recent years, so far we do not have technical solutions. One possible explanation for the lack of progress is a lack of understanding of the problem. We would argue that our understanding of the problem is deficient exactly because we try to understand it as a purely technological challenge. Appreciating algorithmic discrimination as the social-technological problem it is will open new perspectives on how it can be addressed.

This is where we hope to make a contribution in the final part of this manuscript. We believe that the focus on a techno-fix is at least partly due to the particular way in which the problem of algorithmic discrimination is currently framed in the field of NLP. In the next section we will rethink this framing and propose an alternative language to ground work on algorithmic discrimination.



\section{How to move forward}\label{sec:reframing}

\subsection{The delusive lure of the techno-fix}

The observations we made in our case study are not new. Very similar results have been reported, for example, in \cite{blodgett_language_2020}, a study published in 2020. Over the last couple of years various authors have expressed criticism about the focus on quick technological solutions. One might wonder why there is so little change in this direction, while at the same time the number of publications on algorithmic discrimination grew immensely.

There are various reasons for this. Here we want to focus on two misguided assumptions that make the techno-fix so appealing. The first misguided assumption is well known. It is the assumption that technology is inherently value-free, that not only can complex social issues be reduced to a technological one, but that this can be done because of technology’s objectivity and capacity for impartial decisions. This conviction ignores the very real ways in which our tools and technologies reflect and embody the environment they emerge from. These things do not spring from nowhere. Every step of the creation process, from how data is collected to how a tool is implemented and used in the world, contains and expresses underlying societal arrangements. This is not specific to AI technology. \cite{winner_80} showed how certain infrastructure and architectural designs inherently include some and exclude others, thereby embodying pre-existing political dynamics and reinforcing inequality through their own effects. Benjamin \cite{benjamin} makes this point in connection with the topic of this paper by thoroughly describing how racial constructs become materially embodied in technologies, and then contribute to discrimination of minorities. 

Technology is always situated. It emerges from certain contexts and has certain effects, whether intended or not. As such, any technological assessment cannot be simplified to looking at the tool in isolation. Instead, it must be understood as an ecosystem \cite{Stahl22, Stahl23}, as an active element within a larger environment of values and power relations. This is crucial not only for any comprehensive ethical evaluation, but – again --  for the very functioning of democracy. 

We want to add a second misguided assumption here, which is not yet part of the discussion: algorithmic discrimination is generally treated as a {\it safety issue}.  There is an important difference between developing technology that does not contribute to discrimination and, say, integrating safety checks into an automatic door. We can easily agree on what it means for such a door to operate safely. However, as society, we are still struggling with finding a ground-truth when it comes to discrimination. Relying on technocratic solutions or Big Tech to resolve problems like algorithmic discrimination sidesteps valuable democratic processes and hinders societal growth \cite{Weinberg, birhane_algorithmic_2021}. How to achieve respectful, fair and just technologies cannot and should not be decided for us, we, i.e. society, should decide the meaning of these concepts ourselves. After all, once these meanings are formalized into technology, they become reproduced in society at an incredibly large scale. 

Both assumptions lead us to believe that solving algorithmic discrimination is just a matter of fixing the technical system, and that these solutions will benefit society at large. Instead, we need to realize that technology can only be developed in dialogue with society. It needs to be grounded in the academic and societal debate on norms and values and the ethical/political assumptions build into technology need to be transparently presented. 

This is directly related to one of our findings in the case study: the normative/ethical grounding of the research done is still weak.  In the remainder of this section we therefore want to provide some tools that can help with the ethical and normative grounding of research on algorithmic discrimination by looking at the vocabulary presently in use and the implicit value-sensitive choices that have already been made here.



\subsection{The limitations of harm and bias}
\label{sec:injustice}

In our case study, we saw that the terms {\it bias} and {\it harm} heavily dominated the conceptual discussion in the articles we reviewed (see Figure~\ref{fig:norm}). However, the usefulness of these terms in providing solid normative anchors for work on algorithmic discrimination is rather limited.

Let us first look at {\it bias}. Although this term is often used with a normative connotation (bias is bad), it is not an inherently ethical notion. Finding a bias in some system's behavior is not necessarily related to an actual wrongdoing. The ethical grounding still needs to be elaborated on; questions like what is wrong about this bias and why, who is wronged, and how; still need to be answered. Thus, while {\it bias} is an important and useful notion in the debate, it does not provide a normative grounding by itself.
 
{\it Harm} fairs somewhat better.  Using harm allows for an intuitive and immediate grounding of any norms one might want to propose – of course, technology should not do harm; the user should not suffer in any sense. Still, there are also issues with this notion. Some of these problems have been already pointed out by others. For instance, one still needs to define what exactly is meant by harm, who is harmed, and in what situations. In order to understand the problem, make a normative assessment, and develop solutions, this is all important information; just pointing to the potential of harm is not sufficient. In addition, the system behavior under scrutiny needs to be thoroughly related to harm. For example, take the very common claim in our sample that harm is caused by bias in language models. This is often all that is said about what is problematic about model bias. The authors proceed by developing and/or evaluating some measure for this bias. However, pointing to performance on these measures is not equivalent to showing that harm will or will not occur, because the relation between these abstract measures and actual harm is at best unclear.

There are some more fundamental issues with harm that should be considered. Most importantly, the concept of harm may not offer as firm an ethical grounding as initially imagined and, therefore, cannot be used to fully explain what is wrong about discrimination. There is a conceptual difference between harm and wrong \cite{feinberg}. Only the latter possesses moral weight since not all harms are necessarily wrongs. Imagine someone punished for disobeying the law or even someone who is harmed by a coconut falling on their head. These are cases in which an individual is harmed, however whether they have been wronged is not as clear, and this distinction has serious repercussions on questions of moral retributions and responsibilities. A harm becomes a wrong when an action is unjustified, only then does it become morally indefensible, and calls into question the relationships between agents involved and what they owe each other \cite{Diberardino}. One consequence of this is that harm is not sufficient for discrimination to take place. 

But it is not just that the notion of harm is not sufficient for discrimination, it also is not necessary. Discrimination can occur without there being a clear sense of harm done. Take the case of the Kenyan workers who get paid 2\$ per hour to clean toxic content from the output of ChatGPT \cite{Perrigo}. In the local context of Angola this is a well-paid job. Therefore, there is certainly no easy way to argue for economic harm in this case. One could argue that there is psychological harm, depending on the content the workers need to clean up \cite{cleaner}. But even in the case that there is no psychological damage involved in this kind of work, it is still morally problematic because it is unjust. These are practices that grew out of post-colonial structures that exploited human beings and lead to an unfair distribution of opportunities and goods of which the effects still resonate today \cite{Hammer}. 

Only taking psychological harm into account may also deflect attention from the true source of the injustice, that is, interlocking systems of oppression and the unjust social practices which support them. One response might be to remedy the psychological damage with individual band-aids such as ’free’ employee therapy sessions or a complementary yoga retreat from Meta, OpenAI or one of their subcontractors. But this is not the right angle to address the problem. What is causing the harm in this case is not what is causing the injustice. 

This example illustrated that framing negative social impact in terms of harm does not provide a proper viewpoint for adequately addressing the wrongs of these technologies and the unjust practices in which they are embedded. The focus on harm (similar for fairness) is connected to a focus on system-user interaction as the locus of trouble: discrimination takes place from output of the system to the agent using it. This is an individualistic framing of the problem of discrimination, which we also see outside the AI-debate. Discrimination is predominantly framed as interpersonal: one person or agent discriminating against another. The cause of discrimination then is located in the individual beliefs and attitude of a person versus another \cite{Becker, Garcia, Correll}. Similarly, the problem of algorithmic discrimination is limited to the NLP system that incidentally might wrong a user or group of users because of prejudiced ’beliefs’ encoded in its internal states. The similarities even extend to blaming the training data (childhood, education) for the problematic ‘beliefs’ (see Figure~\ref{fig:causes} and the dominance of bias in the data set as cause of system bias). The solution then often lies in merely correcting these biases, again something that clearly showed up in our data (see Figure~\ref{fig:solutions}).

To sum up this part of the discussion, harm is neither necessary nor sufficient for discrimination. Most importantly, the focus on harm leads to an interpersonal understanding of discrimination, and reducing discrimination to the interpersonal level creates blind spots for the structural dimensions of discrimination.
We are not arguing to get rid of the notion of harm altogether. On the contrary, there is an important role for harm to play in the context of the debate on algorithmic discrimination. It is one of the central reasons for what is wrong about discrimination. It is just not the only one. In other words, we should keep the notion of harm as referring to damage or suffering that might be caused to particular groups or individuals, but we need a different, more general and morally consequential notion to ground the discussion of algorithmic discrimination.\footnote{One possible candidate for such a notion is fairness. We noticed, for instance, that approaches in our sample that focused on fairness achieved a more thorough conceptual grounding of their normative claims, because of the rich literature that exists on formalizations of fairness \cite{Narayanan} However, we also observed that articles nearly exclusively focused on these formalizations without considering the actual social processes involved (see also \cite{Taylor}). We see this as a weakness in the fairness approach and believe it can be fortified with the framing we put forward here.}

\subsection{Injustice instead of harm} 

Let us go back to our beginning. The problem we are trying to solve here is algorithmic discrimination. Discrimination is treating groups or individuals differently in a way that is morally wrong. In the Introduction we linked this moral wrongdoing to violations of the foundations of democracy.

It is not making a distinction that is problematic – we have to treat different people differently. It is also not that we harm one group and not the other that is at issue – such a distinction can be justified. What is wrong about discrimination is that it is unjust. This latter notion, injustice, lays at the heart of algorithmic discrimination.

Therefore, we propose to replace the notion of harm with that of injustice. Of course, there are also debates regarding definitions of in/justice, and formalizations may be just as varied and complex as those trying to formalize a notion of bias \cite{Edenburg}. However, tackling the complexity of injustice instead of bias would urge researchers using the notion to dig deeper when providing the conceptual and normative grounding of their research rather than resorting to harm or discrimination full stop.  

Here, we want to propose to use work on injustice by Haslanger \cite{haslanger22, haslanger23, haslanger24}, which naturally incorporates many aspects that came up in our discussion. A just system then, according to Haslanger would be one in which the practices and structures of this system would enable individuals to reflect critically on what they value and to pursue these values. This means that the system should support and ensure the social and material conditions for the development of skills, knowledge and resources that this requires. In other words, a just system supports people’s full democratic participation \cite{haslanger24}. A concern for justice is what grants equal opportunities to groups and individuals, it is what guides true democratic participation and deliberation, and it must be practiced at every step of technological development if we want the scientific community to be an ethical and politically responsible contributor to democratic values.

Haslanger's framework also naturally incorporates the structural and systemic aspects of discrimination that were missing when focusing on the notion of harm. She states that ’structural injustice occurs when the practices that create the structure – the network of positions and relations – (a) distort our understanding of what is valuable or (b) organize us in ways that are unjust/harmful/wrong, e.g., by distributing resources unjustly or violating the principles of democratic equality’ \cite{haslanger22}. Following this, systemic injustice ’occurs when an unjust structure is maintained in a complex system that its self-reinforcing, adaptive, and creates subjects whose identity is shaped to conform to it’ \cite{haslanger22}.

To recognize that structural injustice is a fundamental aspect of discrimination subsequently requires more than removing bias from a model or dataset. Even a technology that operates accurately and does not show biased behavior in its user interactions can still contribute to practices that produce unjust social relations. Let us take facial recognition technologies, which often show biases and discriminate against people of color, women, and, women of color in particular \cite{benjamin, Noble, Gebru, Perkowitz2021Bias, Hill}. These biases have been linked to biased or limited training data of these systems. This, in turn has led to calls for more thorough training of these systems on a greater variety of data. However accurate facial recognition can still have discriminatory effects because it can reinforce an unjust structure as part of an oppressive system. This system organizes and (re-)produces subjects based on certain markers, like gender, race, species, ability and so on, which are politically loaded categories. Even when we succeed in debiasing the technology, the technology is not off the moral hook. Making the dataset a more accurate and balanced representation of people might even increase harm for a minority producing an effective tool to enforce (and legitimize!) an oppressive system \cite{benjamin}. While one might argue that this merely points at a problematic use of the technology being targeted at an oppressed minority, this foregoes the fact that these systems are inherently unjust because they rely on categorization based on facial features including skin color.	

To summarize, harm does not give us sufficient ethical or explanatory power. When a harm is also a wrong, the normative dimension becomes irrefutable. Wrongs are inherently unjustified and demand reflection on our moral obligations and responsibilities to each other. Yet injustice as we have described it so far, points at a distinct normative wrong from the perspective of democratic participation. It lets us position our considerations in concrete democratic values and practices. At the beginning we have mentioned that we ground our normative position with respect to discrimination in adopting democracy as a core value. This means embracing equal rights and equal opportunities for all members of society. Since systemic and structural injustice violate these ideals of equality and democracy, using a concept of injustice when framing the problem of algorithmic discrimination gives us both the space and strength needed to make substantive ethical and political arguments for assessing technology.

\section{Conclusions}
\label{sec:conclusion}

In this article we have been taking a meta-perspective, studying practices within the field of NLP in 2022. The goal was to engage with the critique that the field overly focuses on technological solutions when addressing the problem of algorithmic discrimination. Our case study has confirmed this suspicion. It also points to a possible explanation for this limitation: a too narrow framing of the problem, together with a shallow grounding of its ethical and societal dimensions. In the second part of the paper we wanted to provide some tools that could help with finding a more substantial ethical understanding of what algorithmic discrimination entails. Concretely, we have proposed to replace the focus on harm in the debate with one on injustice. Using one particular way to flesh out what injustice means, building on the work of Haslanger, we showed that this allows a broader perspective on the underlying dynamics leading to algorithmic discrimination. An important part of this is understanding the role technology can play in structural, and even systemic forms of discrimination. This framework also highlight the close relationship between discrimination and impeding core practices of a democratic society.

Looking at the problem of algorithmic discrimination from a systemic point of view also implies seeing the NLP practitioners themselves as embedded in social structures. This can mean paying attention to the well-discussed fact that NLP practitioners have certain demographic features and that this affects the way they set research agendas and build AI systems. This is, however, not the only way structural factors influence their work. NLP researchers are embedded in various power structures that steer their behavior. They face significant pressures: to publish, gain conference acceptance, and secure funding. For those working in industry these demands are even more adamant by an additional layer of institutional and commercial constraints. While these pressures shape the current landscape of NLP research, they should not be perceived as immutable limitations. Instead, we must recognize these power structures and use them as a starting point for a discussion on how they can be challenged. In the end, it is the research community itself that defines what counts as good science -- and that means it also holds the power to redefine it.


\bibliographystyle{ACM-Reference-Format}
\bibliography{AIDproject}


\appendix

\section*{Appendix}

\nociteSample{*}
\bibliographystyleSample{ACM-Reference-Format}
\bibliographySample{Sample}

\newgeometry{top=3cm,left=2cm,right=2cm,bottom=2cm}

\begin{figure*}
    \begin{footnotesize}
    \begin{tabular}{p{5cm}p{11cm}}
         \multicolumn{2}{c}{\bf Table of labels for the first research question: Ethical grounding}\\
        \textsc{Label group} & \textsc{Labels}\\
        \toprule[2pt]
        \multicolumn{2}{l}{\it Paper-level labels} \\
        \midrule[1pt]
        wrongness not discussed
        \vspace*{0,1cm}\\
        \multicolumn{2}{p{16,5cm}}{EXAMPLES: This is a paper-level label, so there are no concrete quotes that are labeled. But to illustrate, a paper in this group would only define the subject (bias) like in this quote from [Sample43] 
        :``We define bias as an inclination or prejudice for or against something, e.g. groups, individuals, concepts and behaviors. The term social bias can be used in two senses: an individual’s bias which is explained by the (social) group the individual belongs to, and bias against (social) groups. The latter is typically the focus of bias studies in NLP ... .''. There is no clear statement that bias is negative; it is just described as an inclination in attitude or opinion.} \\
        \midrule[2pt]
        \multicolumn{2}{l}{\it Quote-level labels} \\
        \midrule[1pt]
        {\bf no ethical vocabulary} & 
        negative impact, undesirable behavior, undesired bias/stereotypes, unsafe behavior, violate the law\\
        \multicolumn{2}{p{16,5cm}}{EXAMPLES: ``Societal biases in NLP has raised increasing attention because large-scale LMs containing societal biases can produce \underline{undesirable} biased expressions and have negative societal impacts on the minorities ...'' [Sample29] 
        , ``In recent years, there has been a series of works aiming to measure \underline{social biases or other unwanted behaviors in NLP}.'' [Sample39] 
        } \rule{0pt}{3ex} \\
        \midrule[1pt]
        {\bf ethical grounding in terms of harm} &
        allocational harm, environmental harm, generate harmful language, harmful bias/stereotypes, representational harm,  unspecific harm/ harmful behavior\\
        \multicolumn{2}{p{16,5cm}}{EXAMPLES: ``Our results hint at how zero-shot generalization may provide some hopeful representation toward minimizing \underline{harm and bias} in these large-scale language models.'' [Sample4] 
        , ``The word representations that are derived from these models (so-called word embeddings) also retain \underline{potentially harmful biases} contained in the data that are used in the training process ... .'' [Sample23] 
        } \rule{0pt}{3ex} \\
        \midrule[1pt]
        {\bf other ethical vocabulary} &
        lack of accountability, lack of transparency, no equitable NLG/NLP, toxic behavior, unfair behavior, untrustworthy AI  \\ 
        \multicolumn{2}{p{16,5cm}}{EXAMPLES: `` language models (LMs) that empower much of the web as we know it are well known to contain biases that promote \underline{structural discrimination} in downstream tasks against minorities and larger social groups alike ... .'' [Sample23] 
        , ``recent studies criticize that the multilingual textual representations do not learn \underline{equally high-quality representations for all the languages} ... .'' [Sample45] 
        } \rule{0pt}{3ex} \\
        \midrule[2pt]
        \multicolumn{2}{l}{\it Second order quote-level labels}\\
        \midrule[1pt]
        {\bf development of concepts} & developed, not developed\\
        \multicolumn{2}{p{16,5cm}}{EXAMPLES: `` If there exists a correlation between a certain gender and an attribute, the language model intrinsically and perpetually causes representational harm \cite{blodgett_language_2020} through improper preconceptions.'' [Sample2] 
        is labeled as developed; ``... LMs propagate stereotypical behavior which could be toxic towards any demographic group (more towards disadvantaged groups).'' [Sample1] 
        is labeled as not developed, because it is not made explicit how propagating stereotypical behavior can be toxic.} \rule{0pt}{3ex} \\
        \midrule[1pt]
        {\bf grounded in theoretical literature} & theoretical grounding, no theoretical grounding\\
        \multicolumn{2}{p{16,5cm}}{EXAMPLES: The first quote from the last examples provides theoretical grounding, the second does not. The criteria here is whether literature is cited that uses theoretical work from social science, political sciences, philosophy or neighboring disciplines.} \rule{0pt}{3ex} \\
        \midrule[2pt]
    \end{tabular}
    \caption{\footnotesize This table gives an overview of the labels used to annotate for the first research question. We distinguish three types of labels: (i) paper-level labels can only given to the paper as a whole, (ii) quote-level labels are given to quotes in papers, consequently, papers can have multiple labels of this type in the same category, (iii) second-order quote-level labels are labels that need to be attached to quote-level labels within that category. 
    }
\label{tab:label_table}
\end{footnotesize}
\end{figure*}


\begin{figure*}
    \begin{footnotesize}
    \begin{tabular}{p{5cm}p{11cm}}
         \multicolumn{2}{c}{\bf Table of labels for the second research question:}\\
         \multicolumn{2}{c}{\bf Causes of wrongful behavior of AI systems}\\
        \textsc{Label group} & \textsc{Labels}\\
        \toprule[2pt]
        \multicolumn{2}{l}{\it Causes of wrongful behavior of AI systems} \\
        \midrule[1pt]
        {\bf issues with training} & 
        biased/limited training data\\
        \multicolumn{2}{p{16,5cm}}{EXAMPLES: "Potential harms arise when biases around word choice or grammatical gender inflections reflect demographic or social biases (Sun et al., 2019).'' [Sample37] 
        .} \rule{0pt}{3ex} \\
        \midrule[1pt]
        {\bf technological causes} &
        bias in AI system, tech fix, impaired performance\\
        \multicolumn{2}{p{16,5cm}}{EXAMPLES: `` If there exists a correlation between a certain gender and an attribute, the language model intrinsically and perpetually causes representational harm (Blodgett et al., 2020) through improper preconceptions.'' [Sample2] 
        is an example for {\it bias in AI system}. ``Differential privacy may often hurt minority groups the most, and reducing the fairness gap by focusing on minority groups during training typically puts their privacy at risk.'' [Sample34] 
         for {\it tech fix}.} \rule{0pt}{3ex} \\
        \midrule[1pt]
        {\bf issues with development} &
        lack of investment in minorities, methodologically weak research, unsupervised automation\\ 
        \multicolumn{2}{p{16,5cm}}{EXAMPLES: ``Languages without large corpora also often face a lack of support in common NLP tools and unexpected errors in other tools due to a lack of testing and use.'' [Sample7] 
        exemplifies {\it lack of investment in minorities.} ``... researchers’ unsystematic approach to experimental design when measuring bias in language generation is a symptom of 1) under-utilizing domain-expert, human annotators to annotate bias generation validation sets and 2) undermining the importance of certain settings that are usually deemed incidental for other NLP tasks, which in fact could be pivotal in bias measurement task.'' [Sample3] 
        was labeled as {\it methodologically weak research}.} \rule{0pt}{3ex} \\
        \midrule[1pt]        
        {\bf issues with use} &
        naive use of biased systems\\
        \multicolumn{2}{p{16,5cm}}{EXAMPLES: ``... naive use of ML methods bears the risk of exacerbating bias'' [Sample48] 
        .} \rule{0pt}{3ex} \\
        \midrule[2pt]
        {\it Causes of bias in AI system}\\
        \midrule[1pt]
        {\bf issues with training} & annotator bias, biased/limited training data\\
        \multicolumn{2}{p{16,5cm}}{EXAMPLES: ``... human-generated data used to build these systems is considered the main source of these biases'' [Sample5] 
        .} \rule{0pt}{3ex} \\
        \midrule[1pt]
        {\bf technological causes} & limited model capacity, loss function, model properties\\
        \multicolumn{2}{p{16,5cm}}{EXAMPLES: ``We reveal the capacity of the model and cross-entropy loss in knowledge distillation have a negative effect on social bias.'' [Sample2] 
        , ``These examples convey that biases can be hidden in the attention mechanism, and thus pose the risk of being recovered in representations and predictions.'' [Sample19] 
        .} \rule{0pt}{3ex} \\
        \midrule[1pt]
        {\bf issues with development} & lack of investment in minorities, subjective decisions, (unsupervised) automation\\
        \multicolumn{2}{p{15cm}}{EXAMPLES: ``We find that biases evaluation is extremely sensitive to different design choices when curating template prompts.'' [Sample6] 
        for {\it subjective decision}. } \rule{0pt}{3ex} \\
        \midrule[2pt]
        {\it Causes of issues with training}\\
        \midrule[1pt]
        {\bf societal bias} & \\
        \multicolumn{2}{p{16,5cm}}{EXAMPLES: ``... these systems learn a wide variety of societal biases embedded in human text ...'' [Sample8] 
        .} \rule{0pt}{3ex} \\
        \midrule[2pt]
    \end{tabular}
    \caption{\footnotesize This table gives an overview of the labels used to annotate for the second research question. For this question we did not use paper-level labels or second-order labels. However, because of the central role bias played in many papers we labeled not only for the cause of wrong, but, if bias was mentioned as such cause, we also labeled the cause of bias. We used for both levels of causes the same grouping of labels: issues with training, technological causes, issues with development. Issues with use could not occur as cause of bias in an AI system. Finally, to give a complete picture of the causal network that was considered, we also labeled causes given for the causes of system bias. There was only one relevant group of causes considered: societal biases.}
\label{tab:label_table}
    \end{footnotesize}
\end{figure*}


\begin{figure*}
    \begin{footnotesize}
    \begin{tabular}{p{5cm}p{11cm}}
        \multicolumn{2}{c}{\bf Table of labels for the third research question: Proposed solutions}\\
        \textsc{Label group} & \textsc{Labels}\\
        \toprule[2pt]
        \multicolumn{2}{l}{\it Quote-level labels} \\
        \midrule[1pt]
        {\bf technological solutions} & 
        technological solutions, + transparent AI, + bias measurement\\
        \multicolumn{2}{p{16,5cm}}{EXAMPLES: ``In this work, we specifically investigate how robust NMT is to language imbalance in tokenizer training.'' [Sample49] 
        . ``Explainability techniques (XAI) can potentially help in discovering and quantifying biases.'' [Sample32] 
        for {\it + transparent AI}, ``It is important to perform a systematic, robust and automated bias analysis to help build equitable NLG systems.'' [Sample1] 
        for {\it + bias measure}. } \rule{0pt}{3ex} \\
        \midrule[1pt]
        {\bf solutions targeting the researchers/developers} &
        diversify approach, create awareness, improve conceptualization, provide dataset statements, improve social/cultural awareness of researchers/developers, study context of use, fix problem before launch, improve annotation, report environmental costs\\
        \multicolumn{2}{p{16,5cm}}{EXAMPLES: ``Extending the KG to other languages as well as data sources could yield a more global view on stereotypes regarding a specific entity.'' [Sample17] 
        for {\it diversify approach}, ``To raise awareness and mitigate stereotypical reflection, we need to understand how biases emerge and how they are reinforced.'' [Sample24] 
        for {\it create awareness}, ``We propose that an artifacts statement should be added to this documentation as a way to contribute to diagnosis (and thus mitigation) of pre-existing bias, which is also one of the goals of data statements.'' for {\it dataset statements}.} \rule{0pt}{3ex} \\
        \midrule[1pt]
        {\bf solutions targeting institutionsor infrastructure} &
        policies/guidelines, research/development agenda\\ 
        \multicolumn{2}{p{16,5cm}}{EXAMPLES: ``Concretely, we should strive to establish robust estimators of bias, clean and curated test sets, and guidelines for their rigorous applications within the (often restrictive) confines of language model APIs to avoid measuring the biases that we introduce in the process of detecting them'' [Sample23] 
        , ``This makes it imperative that associated harms be understood not just for the Western world and with a focus on English language models, but also across languages and cultures.'' [Sample30] 
        . } \rule{0pt}{3ex} \\
        \midrule[1pt]        
        {\bf involving other agents in development} &
        interdisciplinary approach, crowdsourcing, domain experts for fairness, community efforts\\
        \multicolumn{2}{p{16,5cm}}{EXAMPLES: ``At the same time, we argue that addressing biases in language models requires and deserves a concerted community effort ...'' [Sample23] 
        , ``The usage of our fair classifiers in the real world should be carefully monitored with the aid of domain experts.'' [Sample38] 
        .} \rule{0pt}{3ex} \\
        \midrule[1pt]
        {\bf others} & provide guidance for users, advise industry practitioners\\
        \multicolumn{2}{p{16,5cm}}{EXAMPLES: ``Furthermore, there is a lack of research advising on how industry practitioners can effectively and cheaply debias outputs whilst retaining quality, accuracy and realism.'', [Sample13] 
        , ``Understanding the risk of uncontrolled adoption of such automatic tools, careful guidance should be provided in how to adopt them.'' [Sample27] 
        .} \rule{0pt}{3ex} \\
        \midrule[2pt]
        {\it Second order quote-level labels}\\
        \midrule[1pt]
        {\bf target solution} & target solution\\
        \midrule[2pt]
    \end{tabular}
    \caption{\footnotesize This table gives an overview of the labels used to annotate for the third research question. For this question we used quote-level labels and one second order label: we labeled all solutions discussed in the paper and then marked the solution the paper actually worked on in the paper with an additional label. In the group {\it technological solutions} we added additional labels for solutions that target transparent AI and bias measurement.}
    \end{footnotesize}
\label{tab:label_table}
\end{figure*}

\end{document}